\documentclass[conference, 10pt]{IEEEtran}
\IEEEoverridecommandlockouts

\usepackage[a4paper, total={184mm,239mm}]{geometry}

\usepackage{graphicx}
\usepackage{amsmath,amssymb,amsfonts}
\usepackage{xcolor}
\usepackage{hyperref}
\usepackage{float}

\usepackage{caption}
\usepackage{subcaption}
\usepackage{multirow}
\usepackage{tabularx}
\usepackage{makecell}

\usepackage{tikz}

\usepackage[ruled,vlined,linesnumbered]{algorithm2e}

\usepackage{pifont}
\newcommand*\circled[1]{\tikz[baseline=(char.base)]{ \node[shape=circle,fill,inner sep=0.4pt] (char) {\textcolor{white}{#1}};}}

\usepackage[T1]{fontenc}

\newcommand{\SX}[1]{\textcolor{blue}{#1}}

\begin{document}

\title{Dataflow Optimized Reconfigurable Acceleration for FEM-based CFD Simulations}

\author{
    \IEEEauthorblockN{Anastassis Kapetanakis, Aggelos Ferikoglou, George Anagnostopoulos, and Sotirios Xydis}
    \IEEEauthorblockA{
        \textit{Microprocessors and Digital Systems Lab, National Technical University of Athens}, Greece \\
        \{akapetanakis, aferikoglou, geoanagn, sxydis\}@microlab.ntua.gr
    }
    \vspace{-30pt}
    \thanks{This work has been partially funded by the EU Horizon 2022 program under grant agreement No 101096698 REFMAP (\url{https://www.refmap.eu/}).}
}

\maketitle

\begin{abstract}
Computational Fluid Dynamics (CFD) simulations are essential for analyzing and optimizing fluid flows in a wide range of real-world applications. 
These simulations involve approximating the solutions of the Navier-Stokes differential equations using numerical methods, which are highly compute- and memory-intensive due to their need for high-precision iterations. 
In this work, we introduce a high-performance FPGA accelerator specifically designed for numerically solving the Navier-Stokes equations. 
We focus on the Finite Element Method (FEM) due to its ability to accurately model complex geometries and intricate setups typical of real-world applications. 
Our accelerator is implemented using High-Level Synthesis (HLS) on an AMD Alveo U200 FPGA, leveraging the reconfigurability of FPGAs to offer a flexible and adaptable solution.
The proposed solution achieves 7.9$\times$ higher performance than optimized Vitis-HLS implementations and 45\% lower latency with 3.64$\times$ less power compared to a software implementation on a high-end server CPU.
This highlights the potential of our approach to solve Navier-Stokes equations more effectively, paving the way for tackling even more challenging CFD simulations in the future.
\end{abstract}


\linespread{0.95}

\section{Introduction}
\label{sec:introduction}

Computational Fluid Dynamics (CFD) play a vital role in analyzing and optimizing fluid flow in complex systems, enhancing design, performance, and safety across industries such as aerospace~\cite{spalart2016role}, automotive~\cite{kobayashi1992review}, and environmental engineering~\cite{janssen2015validation}.
Constructing physical prototypes for studying these problems is costly and time-consuming, particularly for intricate fluid dynamics scenarios that require elaborate setups and significant resources.
CFD simulations offer a cost-effective and efficient alternative by providing detailed insights and flexibility for design optimization without physical models.
They facilitate virtual testing across diverse conditions and use cases~\cite{koliogeorgi2024auto, raman2018review}, helping to identify potential issues early, mitigate risks, and improve overall design performance~\cite{keyes2013multiphysics}.

CFD simulations involve solving the Navier-Stokes Partial Differential Equations (PDEs), which are fundamental in describing fluid flow behavior~\cite{tsai2018lectures}.
Solving these PDEs analytically is challenging, so numerical methods are used to approximate their solutions.
The most common methods are the Finite Difference Method (FDM) and the Finite Element Method (FEM).
FDM~\cite{liszka1980finite} approximates the derivatives in the Navier-Stokes equations using differences between function values at discrete points on a structured grid, facilitating implementation.
However, FDM's dependence on structured grids restricts its effectiveness for complex geometries and irregular boundaries, making it challenging to handle intricate boundary conditions and maintain stability. 
FEM~\cite{felippa2004introduction} discretizes the domain into small elements and employs interpolation functions to approximate solutions, utilizing unstructured meshes that can adapt to intricate geometries and irregular boundaries.
This flexibility makes FEM especially effective for modeling complex real-world applications, such as flows around irregularly shaped aircraft wing~\cite{vos2002navier}. 
However, it comes with increased implementation complexity and higher computational demands.

Numerically solving the Navier-Stokes equations is time-consuming because of the need for fine grid spacing, numerous time-steps to reach statistical steady-state solutions, and the complexity of the algorithms.
To achieve a practical time-to-solution, it is crucial not only to use efficient numerical modeling but also to leverage the advanced capabilities of modern computational architectures.
General-purpose processors, such as Central Processing Units (CPUs) and Graphics Processing Units (GPUs), are commonly used for numerically solving PDEs. 
Current CPU implementations~\cite{fischer2007nek5000} primarily rely on Matrix-Vector multiplication, utilizing matrix processing libraries to enhance speed.
However, this approach has two major drawbacks that hinder performance and efficiency: i) it requires substantial buffer memory to store the large, sparse matrices, and ii) CPUs find it challenging to leverage data and computation reuse in these matrices~\cite{asgari2020alrescha}.
Recently, GPUs have also been utilized to accelerate CFD simulations~\cite{zhu2024freestencil, SOD2D} due to their ability to manage large-scale data and computations more efficiently than CPUs.
Despite these advantages, GPUs continue to struggle with low energy efficiency, even when processing simpler PDEs on smaller grids~\cite{chen20201}.

Several domain-specific accelerators have been developed to numerically solve PDEs using the FDM method.
Chen et al.~\cite{chen20201} proposed an FDM accelerator for 2D Laplace and Poisson equations on grids up to 128x128, utilizing processing-in-memory (PIM), though it suffers from limited computing precision.
Mu et al.~\cite{mu202129} developed an accelerator for the 2D Laplace equation on a 21x21 grid with dynamic computing precision.
In their subsequent work~\cite{mu2022scalable}, they expanded support to 3D Laplace equations on a 16x16x16 grid without external memory accesses, but both designs are limited to specific grid sizes.
Li et al.~\cite{li2023fdmax} recently introduced FDMAX, an elastic accelerator architecture designed to overcome some of these limitations. 
FDMAX can handle 2D Laplace, Poisson, Heat, and Wave equations with arbitrary grid sizes using 32-bit floating-point precision, offering notable improvements in performance and energy efficiency. 
However, existing solutions target the FDM method and do not address the Navier-Stokes equations.
This implies that adapting these architectures would still encounter major limitations due to the inherent constraints of FDM, particularly in handling complex geometries.
Additionally, these accelerators are based on ASIC designs, which are inflexible, costly, and time-consuming to fabricate~\cite{zahiri2003structured}. Once built, ASICs cannot be modified, making it challenging to adjust key parameters like boundary conditions or grid sizes that are essential for realistic CFD simulations.
Recently, Friebel et al.~\cite{friebel} introduced a FEM-based reconfigurable accelerator.
However, their emphasis on resource-constrained FPGAs restricts performance as they do not take advantage of the capabilities provided by modern cloud FPGA devices.

In this work, we present the first FEM-based reconfigurable accelerator specifically tailored for solving the Navier-Stokes equations.
The proposed accelerator employs FEM for spatial discretization, enabling it to adjust to the intricate geometries and setups typical in practical complex CFD applications. We developed our accelerator architecture using High-Level Synthesis (HLS), a user-friendly approach for programming FPGAs.
Rather than opting for an ASIC-based design, we utilize the reconfigurability of FPGAs, enabling flexible and dynamic hardware adaptation.
This allows for adjustments to different grid sizes or boundary conditions, for instance.
From an architectural standpoint, we introduced tailored source code restructurings that enables HLS to exploit Task Level Parallelism (TLP). To further boost TLP efficiency, memory-aware optimizations are proposed to parallelize off-chip memory transfers to the FPGA's reconfigurable fabric, alongside directive-based HLS micro-architectural optimizations to enhance the accelerator's performance through Initiation Interval minimization.
We thoroughly evaluate the performance of our accelerator by deploying it on an AMD Alveo U200 FPGA, showing that the proposed solution achieves 7.9$\times$ higher performance with respect to optimized Vitis-HLS implementations. In comparison with its software implementation counterpart mapped on a high-end server CPU, we show that the proposed solution delivers 45\% latency gains in end-to-end CFD simulations, while dissipating 3.64$\times$ less power. 

\section{Theoretical Background}
\label{sec:background}

\subsection{Navier-Stokes Equations}
\label{ssec:navier-stokes}

The Navier-Stokes PDEs describe the evolution of a fluid's velocity field over time, influenced by forces like pressure, viscous stresses, and external factors such as gravity~\cite{tsai2018lectures}.
We focus on the 3D compressible Navier-Stokes equations, as detailed in~\cite{SOD2D}, which are described by the following equations:

\vspace{-10pt}
\begin{align}
&\frac{\partial \rho}{\partial t} + \nabla \cdot (\rho \boldsymbol{u}) = 0 \label{eq:continuity}\\
&\frac{\partial (\rho \boldsymbol{u})}{\partial t} + \nabla \cdot (\rho \boldsymbol{u} \otimes \boldsymbol{u}) + \nabla p - \nabla \boldsymbol{\tau} = \boldsymbol{f} \label{eq:momentum}\\
&\frac{\partial E}{\partial t} + \nabla \cdot ((E+p)\boldsymbol{u}) - \nabla \cdot (\boldsymbol{\tau} \cdot \boldsymbol{u}) - \nabla \cdot (\kappa \nabla T) = S \label{eq:energy}
\end{align}

Equations~\ref{eq:continuity},~\ref{eq:momentum}, and~\ref{eq:energy} correspond to the conservation of mass, conservation of momentum, and the conservation of energy, respectively. 
The spatiotemporarily dependent variables to be solved are the \textit{fluid density} ($\rho$), \textit{velocity} ($\boldsymbol{u}$), and \textit{temperature} ($T$).
The \textit{total energy} ($E$) and \textit{pressure} ($p$) are related to these variables through constitutive equations, following the ideal gas law. 
The \textit{viscous stress tensor} ($\boldsymbol{\tau}$) represents the internal frictional forces within the fluid caused by its viscosity.
The terms $\boldsymbol{f}$ and $S$ are the \textit{source terms} of Equations~\ref{eq:momentum}, and~\ref{eq:energy}, accounting for external energy inputs/losses within the system. 
The parameter $\kappa$ is a constant that represents the \textit{fluid's thermal conductivity}.
We solve the equations using the initial and boundary conditions defined by the Taylor-Green Vortex (TGV) problem~\cite{debonis2013solutions, SOD2D}.

\subsection{Numerical Methods}
\label{ssec:numerical-methods}

Since the Navier-Stokes equations cannot be solved analytically, numerical methods are used to approximate the solution. 
Given that $\rho$, $\boldsymbol{u}$, and $T$ are spatiotemporally dependent, we employ the Finite Element Method (FEM) for spatial discretization and the Runge-Kutta Method (RK) to compute the system's time evolution.
In the following paragraphs, we outline the fundamentals of these two methods. 

\textbf{Finite Elements Method.} Consider a convection-diffusion model of the form \( A(x) = 0 \), where \( A(x) = M(x) + C(x) + D(x) \). 
The operators \( M(x) = \frac{\partial x}{\partial t} \), \( C(x) = \nabla \cdot \boldsymbol{f}(x) \), and \( D(x) = -\nabla \cdot (\lambda \nabla x) \), correspond to the \textit{Mass}, \textit{Convection}, and \textit{Diffusion} terms, respectively\footnote{Equations~\ref{eq:continuity},~\ref{eq:momentum}, and~\ref{eq:energy} can be mathematically expressed as Convection-Diffusion models~\cite{FEM_book}.}
For the geometry to be simulated, we assume it is represented by a discretized mesh.
The mesh consists of volume elements defined by vertices and edges, allowing for the representation of complex geometries beyond simple cubes.
The unknown function \( x \) at each node of element \( e \) is represented by the vector \( \boldsymbol{x}^e = [x^e_1 \, \dots \, x^e_n]^T \).
To approximate \( x^e \), a linear combination of \( n \) shape functions \( N_i \) is used, each of which is equal to 1 at its respective node and 0 at all other nodes within the same element.
This leads to the trial function \( x^e = \sum_i x^e_i \cdot N_i \).
The trial function is then substituted into the original differential equation to compute the residual of the differential equation for that element.
The goal of FEM is to find the coefficients \( x^e_i \) for all elements such that:

\vspace{-10pt}
\begin{align}
\label{pre-quadrature}
\sum_e \int_{V_e} N_i \cdot A(x^e)dV = 0 \text{    } i = 1, \dots, n 
\end{align}

Since these integrals are typically not solvable in closed form, the Gauss-Lobatto-Legendre (GLL) numerical integration technique is employed. 
This reformulates Equation \ref{pre-quadrature} as:

\vspace{-10pt}
\begin{align}
\sum_e \sum_g W_g N_i(\boldsymbol{\xi_g}) \cdot A(x^e(\boldsymbol{\xi_g})) &= 0 \text{    } i = 1, \dots, n  \label{eq:fem_implementation}
\end{align}

\noindent where \( W_g \) and \( \boldsymbol{\xi_g} \) denote the quadrature weights and points.
The computation of the \textit{Mass}, \textit{Diffusion}, and \textit{Convection} terms at the quadrature points \( \boldsymbol{\xi_g} \) yields a linear system of equations of the form \( \boldsymbol{K} \boldsymbol{\tilde{x}} = \boldsymbol{b} \) where \( \boldsymbol{\tilde{x}} = [(\boldsymbol{x}^{e_1})^T \dots (\boldsymbol{x}^{e_m})^T]^T \) combines the unknown vectors of all elements. In this linear system, $\boldsymbol{b}$ represents a constant term, while $\boldsymbol{K}$ is a diagonal matrix that encapsulates the information from Equation~\ref{eq:fem_implementation}.

\textbf{Runge-Kutta Method.} Consider an initial value problem for an Ordinary Differential Equation (ODE) of the form $\frac{dy}{dt} = f(t, y), \quad y(t_0) = y_0$.
The Runge-Kutta method~\cite{cartwright1992dynamics} is a numerical approach used to solve ODEs when analytical solutions are difficult or impossible to obtain. 
It enhances simpler methods by offering more accurate approximations at each time step, typically by evaluating the slope at multiple points within the interval.
Following the approach outlined in~\cite{SOD2D}, we employed the fourth-order Runge-Kutta method (RK4), known for its effective balance between accuracy and computational efficiency.

\subsection{Source Code Description \& Characterization}
\label{ssec:pseudocode}

\begin{figure}[t]
    \centering
    \includegraphics[width=\linewidth]{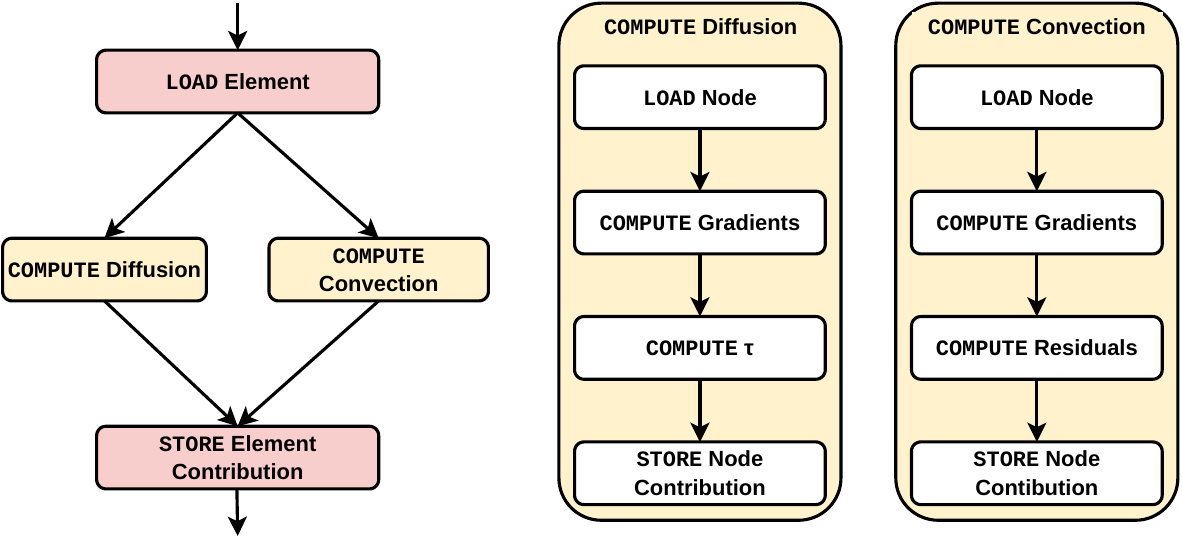}
    \caption{Dataflow Graph of Core Computation}
    \vspace{-10pt}
    \label{fig:algorithm-dfg}
\end{figure}

In this section, we provide a brief overview of the source code developed using the numerical methods discussed in~\ref{ssec:numerical-methods} to solve the problem described in~\ref{ssec:navier-stokes}.
The process begins with loading and preprocessing the discretized mesh.
The main computation takes place within a time-stepping loop, where the algorithm proceeds through four stages of the Runge-Kutta (RK) method at each time step.
During each stage, the algorithm computes the \textit{Diffusion} and \textit{Convection} terms based on FEM. 
As illustrated in Figure~\ref{fig:algorithm-dfg}, the algorithm calculates the contribution of each grid element \(e\) to the diffusion and convection terms by loading the element's data, independently computing both terms, and storing the results for the next iteration.
This process requires computations across the nodes of each element.
First, the node data are retrieved, followed by the computation of the gradient, $\boldsymbol{\tau}$, and residuals.
Finally, the node's contribution is stored.
After each RK step, the algorithm updates the values of $\rho, \boldsymbol{u}, T, E$, and $p$.
This procedure continues until the solution is computed for all time steps.

\begin{figure}
    \centering
    \includegraphics[width=0.65\linewidth]{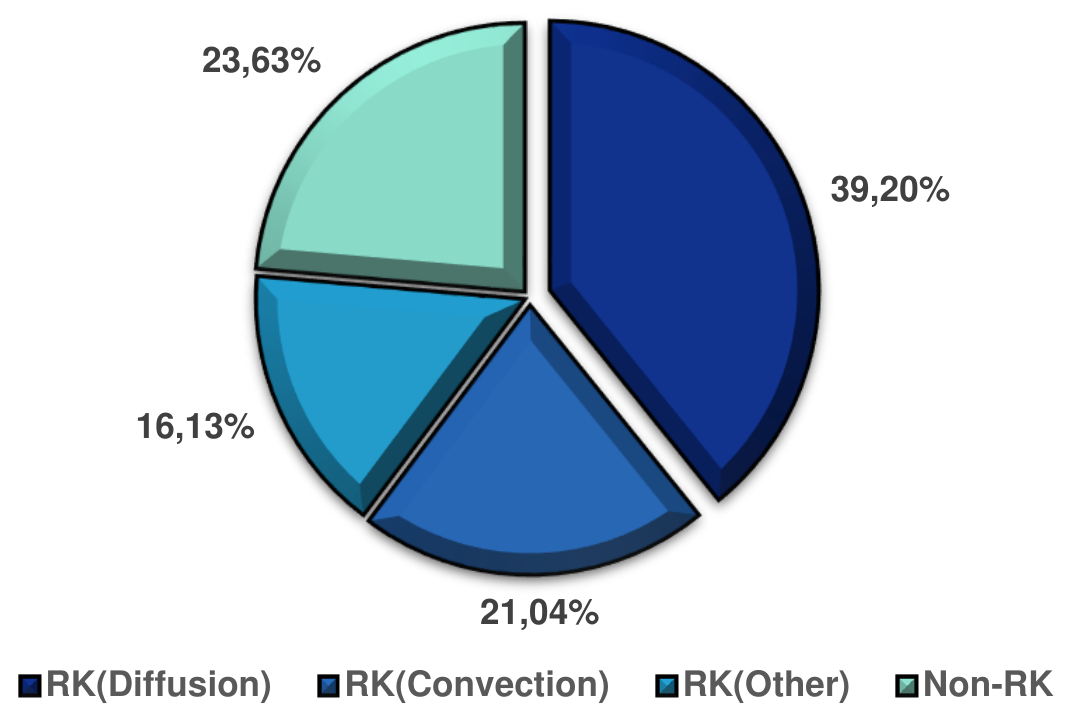}
    \caption{Breakdown of Average Execution Time}
    \vspace{-10pt}
    \label{fig:profiling_data}
\end{figure}

To pinpoint the most time-consuming parts, we performed a detailed profiling analysis across different input sizes, .i.e. number of mesh nodes, ranging from 1M to 4M.
Figure~\ref{fig:profiling_data} shows the average breakdown of execution time.
As shown, the RK method was the most time-intensive, accounting for an average of 76.5\% of the total execution time.
Within the RK method, the diffusion and convection functions emerged as the primary hotspots, consuming 39.2\% and 21.04\% of the total execution time, respectively. 
Therefore, the entire RK method is amenable for acceleration, with particular emphasis on optimizing  convection and diffusion. Similar profiling data have been also recently reported~\cite{koliogeorgi2024auto} targeting FEM-based multi-GPU CFD, further strengthening the validity of our results.

\section{FEM Reconfigurable Accelerator Design}
\label{sec:accelerator-design}

\begin{figure*}[t]
    \centering
    \includegraphics[width=\linewidth]{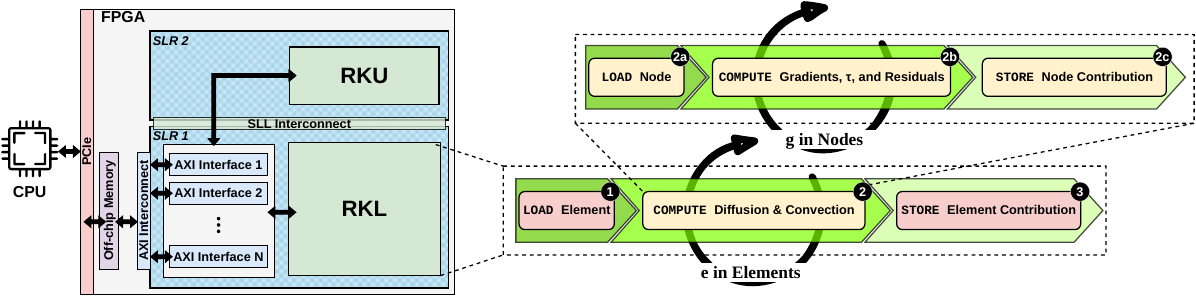}
    \caption{Overview of the Proposed Accelerator's Architecture}
    \vspace{-10pt}
    \label{fig:accelerator-design}
\end{figure*}

The following sections offer a detailed overview of the proposed accelerator architecture and the inter-task, memory and intra-task micro-architectural optimizations introduced for the RK method, i.e. the most computationally demanding component Section~\ref{ssec:pseudocode}. The remaining computations are handled by the host CPU. 
We utilize HLS for introducing the proposed accelerator architecture  and optimizing its FPGA mapping.  

\subsection{Accelerator Architecture}

An overview of the proposed accelerator is illustrated in Figure~\ref{fig:accelerator-design}. 
The CPU is responsible for handling the initialization and overseeing the iteration process across the time steps.
It is also responsible for transferring the necessary data via PCIe to the off-chip memory of the target FPGA.
On the other hand, the accelerator is divided into two separate kernels, with each assigned to a different Super Logic Region (SLR).
The \textit{Runge-Kutta Loop} (RKL) kernel performs the core computations, while the \textit{Runge-Kutta Update} (RKU) kernel evaluates $\rho$, $\boldsymbol{u}$, $T$, $E$, and $p$ at every time step.
RKL is assigned to the SLR with direct access to off-chip memory, while RKU connects to the same memory through the \textit{Super Long Lines} (SLL) interconnect.
Although SLL connections have higher latency compared to intra-SLR fabric connections~\cite{di2023leaps}, RKU is less time-consuming, and its data access patterns are far more regular than those in RKL. This RKL-RKU partitioning enables more effective utilization of FPGA's resources, by de-stressing the Place\&Route phase from trading performance optimality for high resource utilization and routing congestion.


RKL is optimized to effectively carry out the core computations of the FEM-based CFD simulation.
Specifically, for the computation of the \textit{Diffusion} and \textit{Convection} terms, the original source code was reorganized into a \textit{Load}-\textit{Compute}-\textit{Store} form to exploit \textbf{Task Level Pipelining} (TLP).
Initially, the data required for each element is transferred in batches from off-chip memory to the BRAMs and URAMs within the Programmable Logic (PL)~\circled{1}. 
The next step involves executing the computations for the \textit{Diffusion} and \textit{Convection} terms~\circled{2}. 
Since these computations share significant functionality, as shown in Figure~\ref{fig:algorithm-dfg}, and no data dependencies are present, we combined these operations into a single module to improve hardware reuse during each element's computation. 
This stage uses the node data already stored in the PL~\circled{2a}, calculates the gradients of both terms, as well as the $\boldsymbol{\tau}$ and residuals~\circled{2b}, and finally stores the node's contribution to the total diffusion and convection calculations~\circled{2c}. 
Once the diffusion and convection computations for the current element are completed, the data is written back to off-chip memory~\circled{3}.
Since the \textit{Load}, \textit{Compute}, and \textit{Store} tasks are sequential, where each task produces data that the next one consumes, they can be pipelined across iterations to enhance performance.
After completing the RKL computations for the current time step, the RKU evaluates $\rho$, $\boldsymbol{u}$, $T$, $E$, and $p$. 
Once completed, the iteration concludes, and the CPU begins the next one.

\subsection{Task Level Pipelining}
\label{ssec:task-level-pipeling}

Utilizing \textit{instruction-level optimizations}, such as loop pipelining and unrolling, is a standard approach when accelerating applications on FPGAs with High-Level Synthesis.
However, relying solely on these optimizations often results in limited overall performance, resource over-utilization, and memory bandwidth bottlenecks, as they focus primarily on fine-grained loop-level improvements.
For example, applying loop pipelining to the outer loop of a nested structure, which is common in most scientific computations, often requires fully unrolling the inner loops for the pipeline to function effectively, leading to kernels that may exceed the available resources of the FPGA.
Moreover, these approaches do not exploit coarse-grained parallelism, which limits scalability and leaves substantial performance potential unutilized.

\textit{Task Level Pipelining} (TLP) forms a key factor for effective \textit{Dataflow Optimization}, and it is an effective approach for addressing these limitations, providing enhanced performance and improved resource efficiency.
TLP involves partitioning the core computation into $N$ sequential tasks, $Task_1, Task_2, \dots, Task_N$, where each task passes data to the next through inter-task buffers, which can be either First-In-First-Out (FIFO) or Ping-Pong (PIPO) buffers.
The $N$ tasks form the TLP stages, with the most time-consuming task determining the Initiation Interval (II), which represents the number of clock cycles needed before the next iteration of the entire pipeline can begin.
In a specific time step of the pipeline, the inter-task buffers temporarily store the data produced by $Task_k$. The data will then be processed by $Task_{k+1}$, while $Task_k$ runs concurrently the next iteration.

As illustrated in Figure~\ref{fig:accelerator-design}, we identified two areas where TLP can be applied: (a) the element-wise computations (i.e., tasks \circled{1}, \circled{2}, and \circled{3}), and (b) the node-wise computations within each element (i.e., tasks \circled{2a}, \circled{2b}, and \circled{2c}).
Noting that the \textit{Diffusion} and \textit{Convection} terms share considerable functionality and perform nearly identical computations, and with no data dependencies detected, we code-merged these similar operations into a single function/module to enhance hardware reuse.
To apply TLP optimization effectively and prevent deadlocks, two key conditions were considered~\cite{VitisHLS_Manual}. 
First, the \textit{Single-Producer-Single-Consumer} rule was established to ensure that each task has a single producer providing data and a single consumer receiving it, facilitating smooth data flow and preventing conflicts. 
Second, it was ensured that inter-task buffers do not bypass any tasks and transfer data sequentially, thereby maintaining the integrity of the pipeline process.
Meeting these conditions can be particularly challenging for complex applications like our CFD simulation and often requires extensive manual rewriting. 
Sections~\ref{ssec:offchip-memory-transfer-parallelization} and~\ref{ssec:micro-architectural-optimization} provide a detailed examination of how we optimized off-chip memory reads within our tasks, followed by the fine-grained HLS directive optimizations for higher Instruction-Level-Parallelism (ILP).


\subsection{Off-chip Memory Transfer Parallelization}
\label{ssec:offchip-memory-transfer-parallelization}

This section discusses the two primary optimizations applied to the tasks in our pipeline that interact with off-chip memory.
These optimizations aim to enhance memory throughput and prevent contention, which can hinder the performance of TLP.

\begin{figure}
    \centering
    \includegraphics[width=1\linewidth]{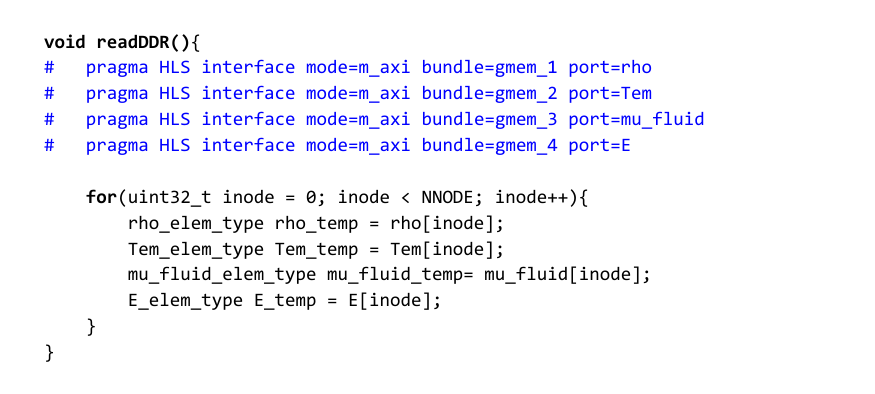}
    \caption{Individual AXI Interface Assignment Optimization}
    \vspace{-10pt}
    \label{fig:axi-bundle-code-snippet}
\end{figure}

\textbf{Arrays to Memory Channel Assignment.} 
The tasks depicted in Figure~\ref{fig:accelerator-design} contain loops that access off-chip memory through one or more AXI interfaces.
These interfaces are connected to an AXI-Interconnect, which in turn connects to the off-chip memory.
Each memory access, typically corresponding to an array element, must be explicitly mapped to a specific AXI interface.
To minimize iteration latency in these loops, we schedule memory accesses concurrently by assigning them to separate AXI interfaces, as depicted in Figure~\ref{fig:axi-bundle-code-snippet}. 
Figure~\ref{fig:axi-bundle-code-snippet} shows a code snippet of \textit{Load-Element} task \circled{1}, demonstrating how the respective AXI interface directives are applied to assign these memory accesses to different AXI interfaces.
This approach eliminates interface contention, which would otherwise force the memory accesses to occur sequentially. 
However, at certain stages of the algorithm, the arrays to be transferred exceed the available AXI interfaces, making it impossible to assign each array individually.
To overcome this limitation, we implement interface reuse for arrays accessed by different tasks during successive steps of the algorithm, such as the \texttt{LOAD-Element} and \texttt{STORE-Element-Contribution} tasks.
Since these loops are not executed in parallel, this method ensures that arrays sharing the same interface do not compete for memory bandwidth, thereby optimizing data transfer efficiency.

\textbf{Decoupling Memory Load and Store Interfaces:} In the RK method, we often encounter loops iterating over arrays stored in off-chip memory, executing operations like \(x[i] \gets f(x[i], y[i])\), where \(f\) is the function applied to arrays \(x\) and \(y\).
These arrays retrieve data from off-chip memory through an AXI interface.
Consequently, the same AXI interface is responsible for reading the values of \(x\) and writing back the updated results. This inter-iteration dependency hinders loop pipelining, ultimately slowing down the overall execution.
To enable pipelined updates, we introduce an additional interface dedicated to \(x\), where one interface handles reading and the other manages writing. 
This approach resolves the inter-iteration dependency, allowing for pipelined memory updates. 

\subsection{HLS Directives for TLP Initiation Interval Optimization}
\label{ssec:micro-architectural-optimization}

As outlined in Section~\ref{ssec:task-level-pipeling}, the Initiation Interval (II) of TLP is determined by the most time-consuming task. 
In this section, we focus on reducing the TLP's II, and thus consequently, the overall execution time of the simulation.
Since our goal is to reduce the TLP II, we focus on optimizing the task with the highest latency in an iterative manner, i.e. the HLS optimization directives are applied each time to the task exposing the highest latency criticality. 
Optimizing all available tasks could result in resource violations due to the limited capacity of the FPGA. 
Moreover, focusing on low-latency tasks would offer minimal performance improvements.  

We prioritize tasks \circled{2a}, \circled{2b}, and \circled{2c} for optimization since they are the most latency-critical.
Furthermore, these tasks gain from processing data stored directly in the PL, where small matrices are housed in the 32KB BRAMs and larger matrices that surpass BRAM capacity are stored in the 288KB URAMs.
The data is fetched through task~\circled{1}.  
More in detail, 
we primarily concentrate on optimizing intra-task micro-architecture, by applying three specific HLS directives: a) loop unrolling, b) loop pipelining, and c) array partitioning. 
After identifying in each step the most latency critical task, we examined HLS directive placement in an intra-task manner. Specifically, for each critical task identified in the current step, 
1. we extract the for-loops with a high trip count and multiple operations in the loop body, 2. we examine potential inter-iteration dependencies and 3. we apply the loop pipelining directive. 
For these large loops, we did not perform unrolling, as this would duplicate the loop body by the factor used, resulting in high resource utilization.
For the for-loops with small trip counts, we completely unrolled them based on the factors allowed by our available resources.
To enable the parallel data accesses required by our directives, we also apply array partitioning with the appropriate factors.
This procedure is repeated until no further optimization could be achieved, either due to unresolved dependencies or resource over-utilization, which would result in lower clock frequencies.



\section{Experimental Evaluation}
\label{sec:experimental-evaluation}

We implement and evaluate the proposed accelerator architecture 
regarding performance, resources utilization, and energy efficiency.
To convert the developed C++ simulation source and its optimizations into HDL, we utilized \textit{Xilinx Vitis HLS 2021.1} and the \textit{Xilinx Vitis Unified Software Platform 2021.1}.
We chose the AMD Alveo U200 as the target FPGA. 
The Alveo U200 card includes 3 Super Logic Regions (SLRs) and 4 DDR memories, each with a capacity of 16GB. 
Communication with the host is enabled via PCIe and the Xilinx Runtime (XRT). 

\subsection{Comparison with Vitis-HLS Optimized Design}

\begin{figure}
    \centering
    \includegraphics[width=0.55\linewidth]{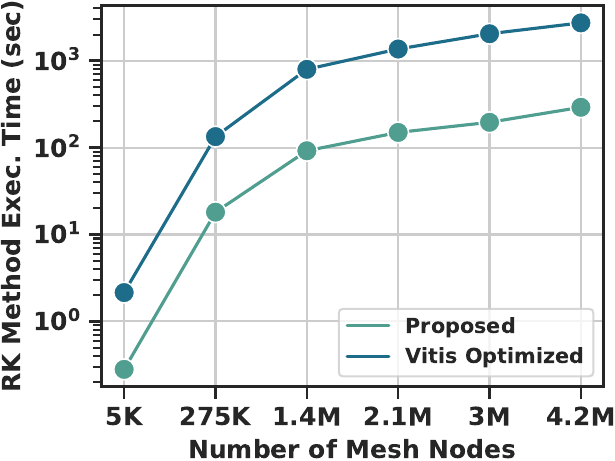}
    \caption{Execution Time for Different Numbers of Mesh Nodes}
    \vspace{-6pt}
    \label{fig:RK_execution_time_sec}
\end{figure}

As an initial step, we compare the proposed accelerator with the Vitis-HLS optimized design. Recent Vitis-HLS release applies the following HLS directive as general optimization strategy: i) automatic loop pipelining using the flag c\textit{onfig\_compile -pipeline\_loops}, ii) unrolling of small tripcount loops through \textit{config\_unroll -tripcount\_threshold}, and iii)
complete partitioning of small arrays using \textit{config\_array\_partition -complete\_threshold}. 
To assess the effect of input data size, we measure the total execution time of the computationally intensive components of our application, specifically RKU and RKL, for varying numbers of mesh nodes.
As illustrated in Figure~\ref{fig:RK_execution_time_sec}, increasing the number of nodes in the examined mesh results in longer execution times for the RK method in both baselines. 
Specifically, increasing the number of nodes from 1.4M to 4.2M results in a $3.4\times$ increase in execution time for both the proposed design and the Vitis-optimized version.
The proposed approach consistently surpasses the Vitis optimization across all tested node counts, achieving an average improvement of $7.9\times$.
The lower performance can be partially attributed to the Vitis-optimized kernel being restricted to a 100 MHz clock frequency, whereas the proposed design operates at 150 MHz.
This limitation  of the Vitis-HLS optimized design arises from both the \textit{RKL} and \textit{RKU} modules being mapped onto the same SLR, which caused significant routing congestion and restricted the maximum clock speed.

Regarding resource utilization, as shown in Table~\ref{tab:utilization}, our optimized design leads to $1.5\times$ higher FF\% and LUT\%, a $1.9\times$ higher BRAM\% and DSP\%, and a $16.8\times$ higher URAM\%, compared to Vitis-HLS optimized design. 
Although the increase in URAM usage is significant, i.e. Vitis-HLS treats URAM as scarce resource, the utilization of other resources shows no more than a two-fold increase compared to Vitis-HLS optimizations.
This indicates that we achieve substantial performance gains with only minimal increases in resource utilization.

\subsection{Comparison with Server CPU}

We compared the proposed FPGA accelerated solution with its software implementation counterpart, i.e. the exact same C++ implementation running in single-threaded mode on a high-performance Linux server, specifically equipped with an Intel Xeon Silver 4210 CPU @ 2.20GHz with 32K L1D/I, 1M L2 and 14M L3 cache.
This choice offers a more balanced comparison than alternatives often cited in the literature, such as the ARM Cortex-A53~\cite{friebel}, due to the Xeon's superior processing capabilities.
For this comparison, we used a 4.2M node mesh, which closely represents a real-world scenario. Our design achieved a 45\% reduction in execution time, demonstrating the effectiveness of the proposed solution.
Another key metric showcasing the potential of our implementation is power consumption.
The CPU implementation consumed an average of 120.42W across all test cases with varying mesh sizes.
In contrast, the FPGA averaged 32.4W for the core application, with an additional 30.7W for peripherals and 1.7W for the rest of the system, resulting in an average power consumption that is $3.64\times$ lower than the CPU.
These initial results highlight the potential of reconfigurable accelerators for efficiently accelerating FEM-based simulations, a field that remains unexplored.

\begin{table}
    \centering
    \begin{tabular}{cccccc} 
          \textbf{Design} & \textbf{FF\%} & \textbf{LUT\%} & \textbf{BRAM\%} & \textbf{URAM\%} & \textbf{DSP\%}\\
          \textbf{Vitis Opt.@100MHz} & 17.19 & 27.68 & 22.96 & 0.73 & 9.17 \\
          \textbf{Proposed@150MHz }& 25.29 & 41.15 & 43.98 & 11.77 & 18.23 \\
    \end{tabular}
    \caption{Post P\&R Resource Utilization Percentages}
    \label{tab:utilization}
\end{table}

\section{Conclusion}
\label{sec:conclusion}

In this work, we presented a high-performance FPGA accelerator tailored for numerically solving the Navier-Stokes equations, with a focus on FEM due to its capability to accurately represent complex geometries and intricate real-world scenarios. 
The proposed accelerator, implemented with HLS on AMD Alveo U200 FPGA, delivers 7.9$\times$ better performance than the Vitis-HLS optimized version, while compared with its software implementation counterpart running on a high-end server it reduces latency by 45\% while consuming 3.64$\times$ less power.
This underscores the potential of our approach for solving the Navier-Stokes equations more efficiently, paving the way for addressing even more challenging CFD simulations.


\bibliography{bibliography.bib} 
\bibliographystyle{ieeetr}

\end{document}